# Chapter 20

# Rare Decays Probing Physics Beyond the Standard Theory

Frederic Teubert

*CERN, Physics Department,*
*Geneva, 23, CH-1211*
*frederic.teubert@cern.ch*

In the last 50 years we have seen how an initially ad-hoc and not widely accepted theory of the strong and electroweak interactions (Standard Theory: ST) has correctly predicted the entire accelerator based experimental observations with incredible accuracy (with the important exception of neutrino oscillation experiments). Decays of the ST particles (quarks and leptons), which are rare due to some symmetry of the theory, have played an important role in the formalism of the ST. These rare decays have been powerful tools to search for new particle interactions with the ST particles, which may not necessarily have the same symmetries. In this article, I will describe the indirect search for evidence of new physics (NP) using quark and lepton flavour changing neutral decays, which are highly suppressed within the ST, and constitute strong probes of potential new flavour structures.

## 1. Historical role of rare decays

Rare decays of mesons have played a very important role in the formalism of the ST of particle physics. In particular, they have been relevant in the development of the Higgs mechanism, which generates fermion masses and quark mixing, and in the establishment of the symmetries of the theory.

A very well known example is the discovery in 1964 that CP was not a symmetry conserved[1] in K decays. When a decade before it had been proposed[2] and then experimentally verified[3] that parity (P) was not conserved in weak decays, nobody imagined that this was also the case for the combination of charge conjugation (C) and P symmetries. It was known by then that flavour was not conserved in weak interactions, and indeed the neutral $K^0$ could mix with its anti-particle[4]. However, if CP was fully conserved, one should be able to define two states ($K_1^0$ and $K_2^0$) that are "eigenstates" of both the weak and CP operators. Subsequently, $K_1^0$ can decay into the pure CP-odd state of two $\pi$ but $K_2^0$ cannot, while at the same time all-possible other decay channels for $K_2^0$ are suppressed by parity violation (semi-leptonic) or by phase-space. Consequently, $K_2^0$ has a much longer lifetime than $K_1^0$ by a factor ~500. Experimenters in Ref.(1) shot protons at a target to produce $K^0$ and after a long enough trip in a vacuum pipe they achieved a pure $K_2^0$ beam. Amazingly, it was measured that very rarely, once in every 500 $K_2^0$ decays, they decayed into two $\pi$ rather than three. At that time only three quark flavours were known, so the observation of this rare $K_2^0$ decay not only implied that CP was not conserved but also that it could not possibly be explained within the theoretical framework currently being used.

Another good example of the influence of rare decays is from around the same time, at the birth of the ST. The weak coupling did not seem to be universal: the observed decay probability of the semileptonic $\pi^+\to\mu^+\nu$ decay, after correcting for the different phase space, was about 20 times larger than the very similar $K^+\to\mu^+\nu$ decay. In 1963 Cabibbo[5] explained these observations by introducing the "Cabibbo angle" ($\theta_c$), such that the weak bosons couple to a linear combination of the *"d"* and *"s"* quarks (using the modern language as in 1963 quarks had not yet been proposed). The suppression of the semileptonic $K^+$ decay with respect to the $\pi^+$ decay arises from the square of the ratio of the linear coefficients, $(\sin\theta_c/\cos\theta_c)^2$, in the decay amplitudes. One would then expect large flavour changing neutral currents (FCNC) in the equivalent case of leptonic $K^0$ decays. However, experimentally the probability of the process $K_2^0\to\mu^+\mu^-$ was measured to be $\sim 7\text{x}10^{-9}$, a very rare decay indeed. This large suppression of observed FCNC motivated Glashow, Ilioupoulos and Maiani[6] in 1970 to predict the existence of an unobserved fourth quark (*c*-quark), which should form an SU(2) doublet with the known *s*-quark. In this model, that later developed into the ST, the existence of two SU(2) doublets, ((*u,d*) and (*c,s*)), implied a series of cancellations which resulted in a strong suppression of FCNC (GIM mechanism), in agreement with experimental observations. There are many other examples of how rare decays of the known existing particles have helped to shape the ST. I've taken these two examples as they demonstrate two important and complementary cases:

- Processes that are protected by some symmetry of the current theory but is not necessarily conserved in the next version of the theory.
- Processes that should not be rare in the current theory, but contradictory observations suggest a new mechanism, which should be included in the next version of the theory.

We will see in the next sections how this historically successful strategy is being pursued today, and what some of the experimental prospects are which can reveal the properties of a theory to supersede the ST. But before that, let's have a look at the flavour structure of the ST and what type of rare decays are *a priori* more interesting to look for.

## 2. Flavour structure and symmetries in the ST

According to the ST the basic constituents of matter can be grouped into three families, or flavours, of quarks and leptons. The four fermions within each family have different combinations of strong, weak, and electromagnetic charges, which determine completely their fundamental interactions with the exception of gravity, which is not included in the ST. As far as we know, quarks and leptons of the second and third family are identical copies of those in the first family but with different, heavier, masses. One of the biggest questions that is not answered by the ST is why do we have three almost identical replicas of quarks and leptons, and what is the origin of their different masses?

In the limit of unbroken electroweak symmetry none of the basic constituents of matter have a non-vanishing mass. The problem of quark and lepton masses is therefore intimately related to one of another open key questions in particle physics: what is the mechanism behind the breaking of the electroweak symmetry? Within the ST these two problems are both addressed by the Higgs mechanism: the masses of quarks and leptons, as well as the masses of the W and Z bosons, are the result of the interaction of these basic fields with a new type of field, the Higgs scalar field, whose ground state spontaneously breaks the electroweak symmetry.

The recent observation by the ATLAS[7] and CMS[8] experiments of a new state compatible with the properties of the Higgs boson (or the spin-0 excitation of the Higgs field) has significantly reinforced the evidence in favor of the Higgs mechanism and the validity of the ST. However, we also have clear empirical indications that this theory is not complete: the phenomenon of neutrino oscillations[9] and the evidence for dark matter[10] cannot be explained within the ST.

The description of quark and lepton masses in terms of the Higgs mechanism is particularly unsatisfactory since the corresponding interactions between fermions and Higgs field are not controlled by any symmetry principle, contrary to all other known interactions, resulting in a large number of free parameters. Besides determining quark masses, the interaction of the quarks with the Higgs is responsible for the peculiar pattern of mixing of the various families of quarks under weak interactions, and of the non-zero CP asymmetry in the ST.

In particular, the interplay of weak and Higgs interactions implies that FCNC processes can occur only at higher orders in the electroweak interactions and are strongly suppressed, in accordance with the GIM mechanism described in Section 1. This strong suppression makes FCNC processes natural candidates to search for physics beyond the ST. If the new degrees of freedom do not have the same flavour structure of the quark/lepton-Higgs interaction present in the ST, then they could contribute to FCNC processes at a comparable level to the ST amplitudes. Even if their masses are well above the electroweak scale, they can produce sizable deviations from the ST predictions for these rare processes.

However, it should be clear from the previous discussion that not all processes that are rare and occur only at higher orders are necessarily powerful tools to look for NP. For instance, within the ST, the interactions between photons and leptons, and between quarks and gluons, are governed by gauge theories (QED and QCD) with non-broken gauge symmetries. The *decoupling theorem*[11] implies that contributions of heavy particles (with masses much larger than the momentum transfer of the process) are irrelevant. This is the reason, for instance, that to a good approximation one does not need to know the values of the top quark and the Higgs boson masses to calculate the value of the running QED coupling constant up to the Z mass. On the other hand, within the ST the weak interactions are described by a gauge theory with broken symmetry. Therefore, higher order corrections are sensitive to the size of the squared mass difference within an isospin doublet (which is a measure of how badly the isospin symmetry is broken). This is the reason why the dominant effect from weak loop corrections within the ST is proportional to ($m^2_{top} - m^2_b$).

Therefore, in general, rare processes that occur at higher orders in the ST are good tools to search for new heavier particles that can modify the loop contribution if they are protected by some symmetry in the ST that does not necessarily hold for the new theory. One example has already been given above, FCNC in quark transitions. Another excellent example is lepton flavour violating decays (LFV). Within the ST neutrinos are massless and the lepton Yukawa matrices can be diagonalized independently. Therefore, there are not FCNC in lepton decays and lepton flavour is conserved. However, the discovery that different neutrino flavours can mix, and therefore that neutrinos have non-zero masses cannot be explained by the ST. Depending on what is the mechanism to generate neutrino masses and therefore what is their nature (Majorana or Dirac) one can expect different levels of charged LFV decays. Searches for LFV decays are therefore extremely interesting, not only as evidence for NP, but as critical data to constrain the mass generation mechanism in the lepton sector, which may or may not be related to the ST Higgs mechanism.

In the next sections I will discuss the status of some of the most interesting examples of rare decays in quark and lepton FCNC. These correspond to the first class of interesting processes described in Section 1. FCNC processes are analogous to the historical discovery of CP violation by looking at decays of $K_2^0 \rightarrow \pi^+\pi^-$, which occur because the symmetry protecting them did not hold in the updated theory.

I also mentioned in Section 1 an historical example of a second class of interesting rare processes that should not have been rare: $K_2 \rightarrow \mu^+\mu^-$. I cannot resist the temptation to just briefly mention an excellent example of this kind of process that is currently being pursued: the search for nucleon electric dipole moments (EDM). There is no good reason why within the ST there is only CP violation in the weak interactions. In general one could add a CP violating term in the Lagrangian describing strong interactions and remain consistent with the ST gauge symmetries. Either there is a new symmetry[12] involving new particles (axion like) and/or a precise measurement of the neutron and proton EDMs will reveal that CP violation is also present in strong interactions at some level.

## 3. Quark flavour changing neutral decays

Within the ST quarks can change flavour in electrically neutral processes via higher order loops. In B-meson decays experimenters have measured *b→s* and *b→d*, in D-meson decays *c→u* and in K-meson decays *s→d* quark transitions. At first order these transitions can occur through two kinds of Feynman diagrams shown in Figure 1. The first corresponds to the so-called "box" diagram and is in particular relevant to describe the mixing between neutral mesons, the example of Figure 1 shows $B_s^0$ mixing. The second kind of diagram, the so-called "penguin" diagram, is responsible for a large variety of FCNC rare decays. The example shown in Figure 1 is that of a *b→sll* transition. In particular, if the bosons radiated are the electroweak bosons (Z, W or γ like in Figure 1), the uncertainties in the calculation of the rates due to less well known non-perturbative QCD effects are drastically reduced as compared with the case where a gluon is radiated. These "electroweak penguins" are particularly interesting for the discussion in this chapter. Let's have a look at today's status of a few interesting examples in the next sections.

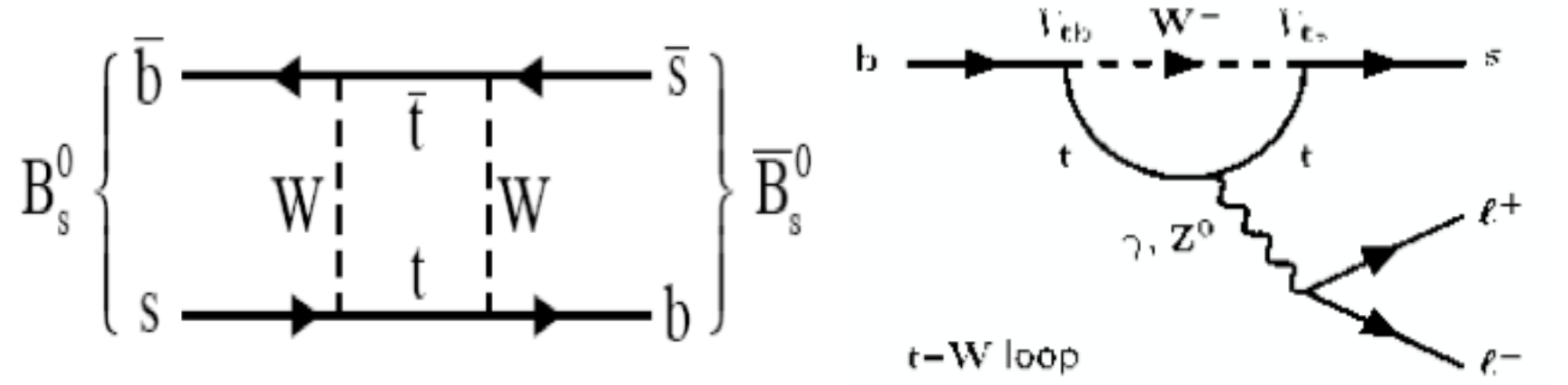

Figure 1 Examples of loop processes within the ST that allow the FCNC *b→s* quark transition. On the left side is an example of a "box" diagram and on the right side an example of a "penguin" diagram.

### 3.1. $K^+\to\pi^+\nu\nu$, $K_L^0\to\pi^0\nu\nu$

One of the strongest constraints on the possible size of NP contributions comes from K physics, in particular the precise measurement of the mass difference ($\Delta m_K$ = $m(K_L)$-$m(K_S)$) of the neutral kaon weak eigenstates and the CP-violating quantities $\varepsilon_K$ and ε'. This is because the ST suppression factors are bigger in the Kaon sector, since the *u* and *c*-quark contributions to FCNC processes are very strongly suppressed by the GIM mechanism, while that of the *t*-quark is strongly suppressed by the Cabbibo-Kobayashi-Maskawa (CKM) matrix elements. Progress in this area is limited by theoretical uncertainties affecting, in particular, the ST prediction of ε'. The situation is better when the process occurs through an "electroweak penguin" with a charged lepton pair in the final state. However, there is still a limitation due to long distance contributions via one or two photon conversions. That's the reason why there is great interest in decays with a neutrino pair in the final state. The $K^+\to\pi^+\nu\nu$ and $K_L^0\to\pi^0\nu\nu$ decays are determined by short distance physics. There is a single operator that determines the decay rates within the ST and in most NP scenarios. In Figure 2 one can see the leading ST Feynman diagrams that contribute to these processes.

Within the ST, these decays are predicted with good precision[13]:

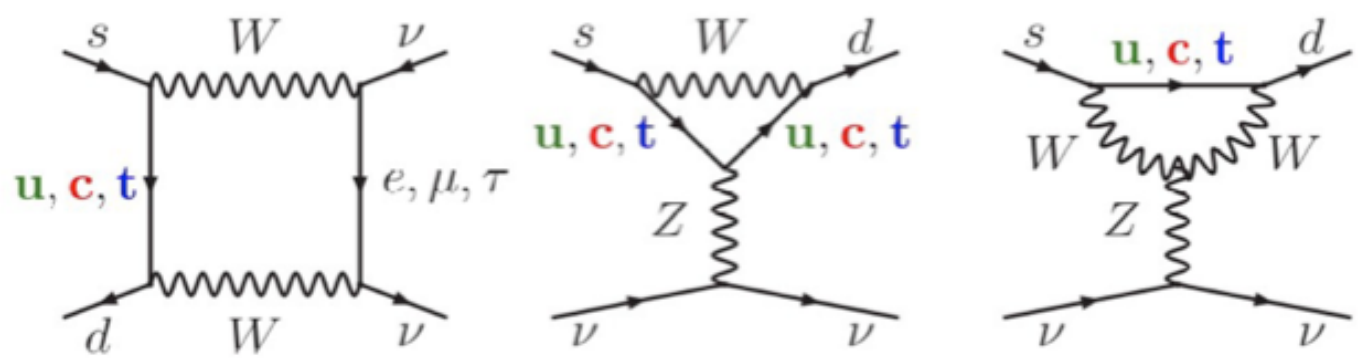

Figure 2 "Box" and "penguin" Feynman diagrams representing the lowest order contributions to the FCNC $s\to d$ quark transition.

$BR(K_L^0 \rightarrow \pi^0 \nu\nu)_{ST} = (2.43 \pm 0.39 \pm 0.06) \times 10^{-11}$
$BR(K^+ \rightarrow \pi^+ \nu\nu)_{ST} = (7.81 \pm 0.75 \pm 0.29) \times 10^{-11}$

where the first uncertainty is due to the experimental uncertainty on the input parameters while the second is due to the intrinsic theoretical limitations. In Figure 3 one can see the results of the experimental searches for these very rare decays in the last 50 years since the birth of the ST. The upgraded E949 experiment at the Brookhaven National Laboratory Alternating Gradient Synchrotron (AGS) achieved the first evidence for the charged K decays with a measurement[14]: $BR(K^+ \rightarrow \pi^+ \nu\nu) = (17 \pm 11) \times 10^{-11}$ while the E391a experiment at KEK in Japan achieved a limit[15] of $BR(K_L^0 \rightarrow \pi^0 \nu\nu) < 2.6 \times 10^{-8}$ at 90% C.L. As accelerator and detector technology has advanced, the sensitivity of rare K decay experiments has also improved. Today, there are two experiments taking data and aiming for a precise measurement of these decays: KOTO at J-PARC in Japan, and NA62 at the SPS at CERN.

The NA62 experiment[16] has the potential to measure the $BR(K^+ \rightarrow \pi^+ \nu\nu)$ with at least a 10% precision. With an expected signal acceptance of ~10% and S/B>4.5, the experiment requires ~$10^{13}$ K decays to achieve such goal. The CERN SPS provides $10^{12}$ 400 GeV protons on target per second, which produces a very high intensity K beam, resulting in 5 million K decays per second in a 60m long vacuum chamber. The sample available to the NA62 experimenters corresponds to ~$4.5 \times 10^{12}$ K decays whose flight path is in their acceptance per year (~$10^7$sec). They expect to see ~45 ST signal candidates per year with <10 background events.

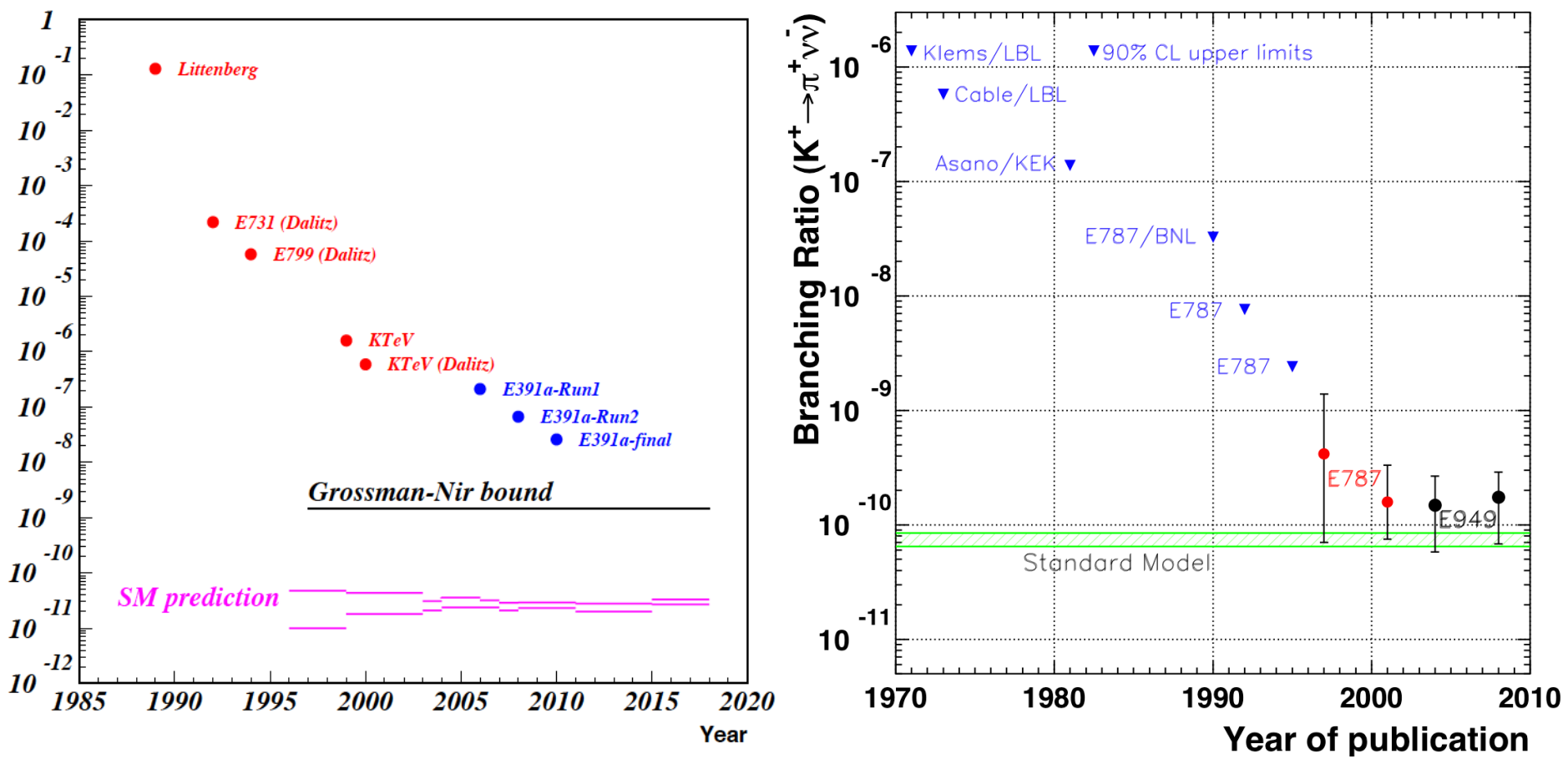

Figure 3 Historical evolution of the 90% C.L. limits and measurements of the branching ratio of the decays $K_L^0 \rightarrow \pi^0 \nu\nu$ (left panel) and $K^+ \rightarrow \pi^+ \nu\nu$ (right panel).

The key for the experiment's success is the background rejection and the uncertainty in the background estimation. The background rejection requires a precise measurement of the incoming K (achieved with a silicon pixel tracker operating at the secondary beam rate of 750 MHz, so-called Gigatracker) and a precise measurement of the outgoing $\pi^+$ as well as a very efficient $\pi^0$ veto. NA62 started commissioning parts of the detector at the end of 2014 and expects to start data taking in 2015.

The KOTO experiment[17] has the potential to reach a first observation of the decay $K_L^0 \to \pi^0 \nu\nu$ at the level of the ST prediction, and has plans for upgrades that could allow for a ~10% measurement of the branching fraction. The J-PARC accelerator can provide $2\times10^{14}$ 30 GeV protons on target every three seconds. Moreover, the neutral K beam is highly collimated ("*pencil beam*") so that the reconstructed $\pi^0$ momentum component transverse to the beam direction can be used as a constraint. The KOTO detector situated ~20m from the target at an angle of 16° from the incident proton beam, has taken data in 2013 for a short period of time (only 100 hours) at ~10% of the nominal intensity. KOTO has been able to observe[18] $\sim 8\times10^7$ K decays reaching a single event sensitivity of $1.3\times10^{-8}$ very similar to the previous best experiment[15]. On the other hand the level of background extrapolated to the signal region is higher than originally anticipated[17] due to a significantly higher contribution from "*halo neutrons*". One signal candidate event is observed compatible with the 0.4 events expected from background. KOTO is expected to resume data taking in 2015.

The $K^+ \to \pi^+ \nu\nu$ and $K_L^0 \to \pi^0 \nu\nu$ decays are indeed extremely interesting measurements in terms of sensitivity to NP. Deviations from the ST at the (10-20)% level[19] appear in many reasonable NP models, compatible with experimental constraints elsewhere. The next decade promises to be very interesting in terms of the potential experimental sensitivity.

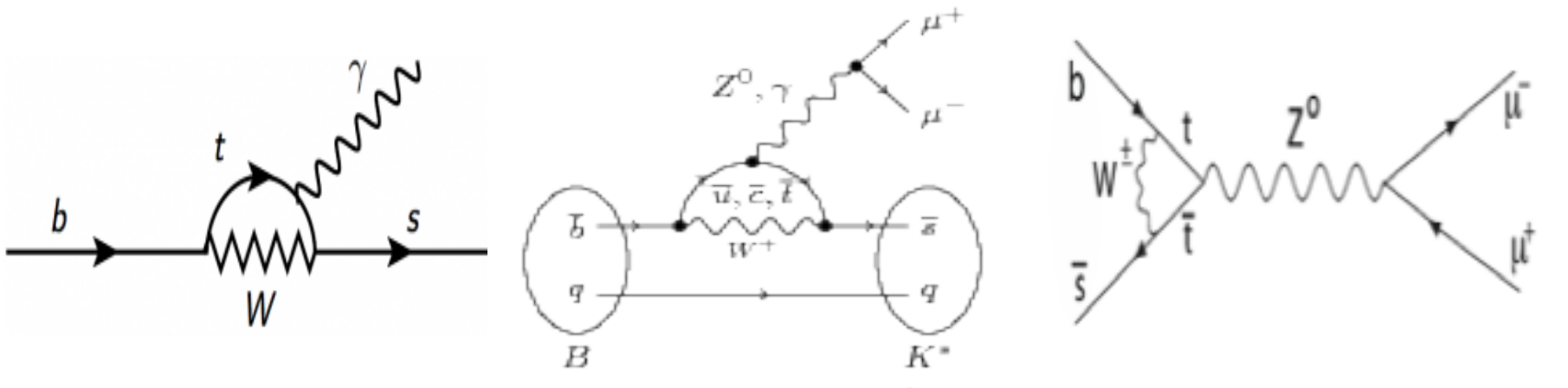

Figure 4 Three realizations of the EW penguin in b→s FCNC. On the left panel one of the simplest form of radiative decays. On the center, the dominant contribution within the ST to the decays $B_d^0 \to K^* \mu^+ \mu^-$. On the right panel the dominant contribution within the ST to the very rare decay $B_s^0 \to \mu^+ \mu^-$.

### 3.2. $B_d^0 \to K^{*0}\mu^+\mu^-$

An example of the simplest realization of the EW penguin in *b→s* FCNC can be seen on the left panel of Figure 4 and consist of a photon emitted from the internal loop. The inclusive process $b \to s\gamma$ has been measured[20] precisely at experiments at KEKB and SLAC, and experiments at CESR and LEP with an uncertainty of ~7%, in agreement with the ST prediction[21]. In fact, this agreement is one of the strongest constraints in NP models (in particular supersymmetric models).

If the photon emitted from the internal loop decays into a lepton pair (hence the amplitude is further suppressed by a factor $\alpha_{QED}$) or it is replaced by a Z boson, the process can provide a rich laboratory to test NP models. An example is shown on the center panel of Figure 4. If we use the language of effective field theories to parameterize NP contributions in terms of a sum of local four-fermion operators ($Q_i$ which depend only on ST fermions) modulated by Wilson coefficients ($C_i$ which depend of the heavy degrees of freedom, i.e. NP particles), then $B_d^0 \to K^{*0}\mu^+\mu^-$ is the "golden mode" to test new vector (axial-vector) couplings (i.e. $C_9$ and $C_{10}$ in terms of Wilson coefficients contributing to the *b→s* transition). Incidentally, $B_d^0 \to K^{*0}\mu^+\mu^-$ complements the $b \to s\gamma$ decay which is mostly sensitive to NP dipole operators (i.e. $C_7$) and the $B^0_{(d,s)} \to \mu^+\mu^-$ decays mostly sensitive to NP (pseudo-) scalar operators (i.e. $C_S$, $C_P$). The charge of the pion in the decay $K^* \to K\pi$ allows the flavour of the B meson to be established; hence an angular analysis can be unambiguously performed to test the helicity structure of the "electroweak penguin".

The system is completely defined by four variables: $q^2$, the square of the invariant mass of the dimuon system, $\theta_l$, the angle between the positive lepton and the direction opposite the B-meson in the dimuon rest frame, $\theta_K$, the equivalent angle of the $K^+$ in the $K^*$ rest frame and $\phi$ the angle between the two planes defined by $(K,\pi)$ and $(\mu^+,\mu^-)$ in the B-meson rest frame. The four fold differential distribution contains a total of eleven angular terms that can be written in terms of seven $q^2$ dependent complex decay amplitudes. These amplitudes can be expressed in terms of five complex Wilson coefficients ($C_S$, $C_P$, $C_7$, $C_9$ and $C_{10}$), their five helicity counter-parts and six form-factors, which play the role of nuisance parameters in the fit.

The LHCb experiment at the LHC is designed to profit from the enormous production rate of *b*-quarks (~$3\times10^{11}$ per $fb^{-1}$) and *c*-quarks (~$6\times10^{12}$ per $fb^{-1}$) in proton-proton collisions at the LHC energies (7-8 TeV of Run-I) in the forward region. About 40% of these *b*-quarks hadronize to form a $B_d^0$ meson and about one in $10^6$ decay into $B_d^0 \to K^* \mu^+ \mu^-$. With ~3 $fb^{-1}$ of data collected in 2011-2012, LHCb has been able to trigger and select ~2400 candidates in the range 0.1 $GeV^2$ < $q^2$ < 19 $GeV^2$ with S/B>5. This is about one order of magnitude larger than the samples available at previous experiments (BaBar, Belle and CDF) and similar to what the experiments ATLAS and CMS, also at the LHC, have collected with ten times more luminosity (however ATLAS and CMS with significantly worse S/B).

The statistics and the quality of the data accumulated by the LHCb experiment allows for the first time to perform a full angular analysis of these decays. The preliminary results of this "tour de force" analysis have been recently shown[22] at conferences. In Figure 5 two examples of the CP-averaged (i.e. the average of the coefficients measured with $B_d^0$ and anti-$B_d^0$ decays) angular coefficients measured by LHCb are shown as a function of $q^2$. While most of the measurements agree reasonably well with the ST predictions[23], for these two examples: $A_{FB}$ (modulating the $\sin^2\theta_K \times \cos\theta_l$ angular term) and $S_5$ (modulating the $\sin(2\theta_K) \times \sin\theta_l \times \cos\phi$ angular term) there seems to be a hint of disagreement. These are early times and more data and a careful reassessment of the ST prediction uncertainties are needed before reaching a conclusion.

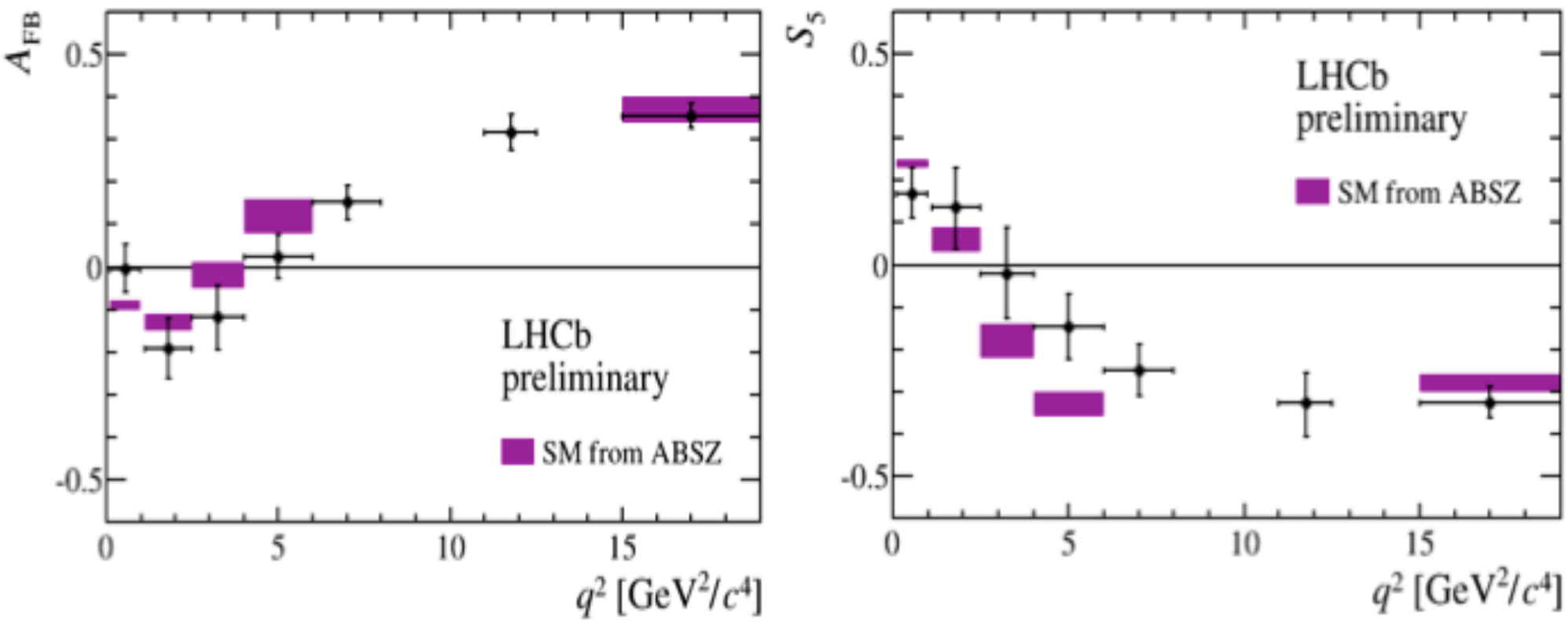

Figure 5 Two examples of the CP-averaged coefficients in the angular terms as a function of $q^2$. The shaded boxes show the ST prediction taken from Ref.(23).

Nevertheless, several authors have already attempted to see if the overall pattern of the measurements is consistent with a given value for the relevant Wilson coefficients. As mentioned before the inclusive *b→sγ* measurements strongly constrain non-ST values for $C_7$. As will be described in the next section the scalar $C_S$ and pseudo-scalar $C_P$ coefficients are constrained, for example, by the measurement of the branching ratio of the very rare decays $B^0_{(d,s)} \to \mu^+\mu^-$. Therefore, the small disagreements observed in the full angular analysis of the decay $B_d^0 \to K^*\mu^+\mu^-$ and also other similar decays like $B_s^0 \to \phi\mu^+\mu^-$ or $B^+ \to K^+\mu^+\mu^-$ seem to be consistent with a non-ST value of the $C_9$ Wilson coefficient, as can be seen in Figure 6 taken from Ref. (24). All these measurements are bound to significantly improve in the next decade. Therefore if NP is the responsible for these early hints, we should get a definitive answer in the near future.

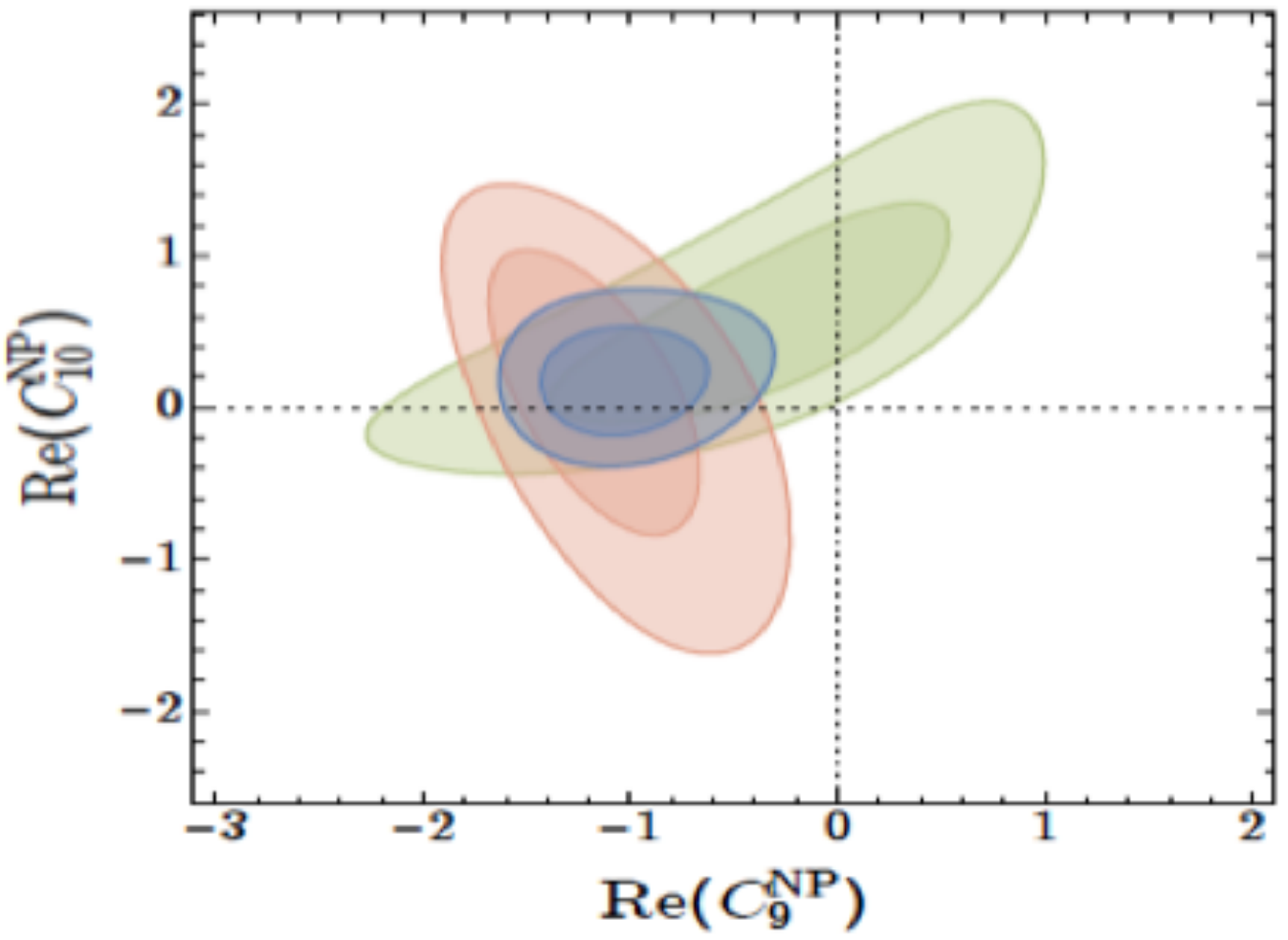

Figure 6 Constrains on the contribution of NP to the real part of $C_9$ and $C_{10}$ at the $1\sigma$ and $2\sigma$ level taken from Ref. (24). The blue contours correspond to the global fit. The red contours correspond to the results using only the full angular analysis of the decay $B_d^0 \to K^*\mu\mu$. The green contours are from other measurements. By definition the ST prediction is (0,0).

### 3.3. $B^0_{(d,s)} \to \mu^+\mu^-$

The pure leptonic decays of K, D and B-mesons are a particular interesting case of "electroweak penguins", see right panel of Figure 4. Compared with the decays described in Section 3.2, the helicity configuration of the final state suppresses the vector (axial-vector) contribution by a factor proportional to $(M\mu/M_{K,D,B})^2$. Therefore, these decays are particularly sensitive to new (pseudo-) scalar interactions. In the case of $B_d^0$ and $B_s^0$-meson decays the contribution of the absorptive part can be safely neglected. As a consequence, the rate is well predicted[25] theoretically: $BR(B_s^0 \to \mu+\mu-)=(3.65\pm0.23)\times10^{-9}$ and $BR(B_d^0 \to \mu+\mu-)=(1.06\pm0.09)\times10^{-10}$. In the $B_s^0$ case, this prediction corresponds to a flavour-averaged time-integrated measurement, taking into account the correction due to the non-vanishing width difference.

The experimental signature is sufficiently clean to reach an expected $S_{ST}/B\sim3$ for the $B_s^0$ decay. The main background in the invariant mass region around the $B_s^0$ mass is due to combinations of real muons and can be estimated from the sidebands. The most important handle to reduce this combinatorial background is the invariant mass resolution of the experiments, which is also crucial to differentiate between the $B_d^0$ and $B_s^0$ decays ($\Delta m\sim87$ MeV). Moreover, the large fraction of $B^0_{(d,s)} \to hh$ decays is an important source of background due to misidentified hadrons in the region around the $B_d^0$ mass (very small in the $B_s^0$ mass region). Given the experimental detector resolution and trigger acceptance, the CMS experiment with 25 $fb^{-1}$ of data accumulated at RUN-I has similar sensitivity to the LHCb experiment with $3fb^{-1}$ collected in the same period. Both experiments[26] have seen a clear observation of the decay $B_s^0 \to \mu^+\mu^-$ (CMS and LHCb observe 28 and 11 $B_s^0$ signal candidates with a background of ~10 and 3.6 respectively) but not yet a significant observation for the decay $B_d^0 \to \mu^+\mu^-$. The combined analysis of the two experiments has been published recently in Nature[27]. The invariant mass distribution can be seen in Figure 7 and the combined value for the branching ratio is measured to be $BR(B_s^0 \to \mu^+\mu^-)=(2.8+0.7-0.6)\times10^{-9}$ in agreement with the ST within the present uncertainties.

It is a testimony to the experimenters ingenuity and persistence that after ~30 years of searching for this very rare decay, since the discovery of B mesons, it has been finally measured at the LHC and found to be in agreement with the ST within the current uncertainties. In the next decade it will be very interesting to see how it turns out the measurement of $BR(B_d^0 \rightarrow \mu^+\mu^-)$.

## 4. Lepton flavour changing neutral currents

The search for FCNC in charged lepton decays has been unsuccessful so far[28], as can be seen in Figure 8. Nevertheless, historically these limits have played a very important role. For example, by the end of the 50s it was known that $BR(\mu \rightarrow e\gamma)$ had to be below $\sim 10^{-5}$ and this was used to argue the existence of a second neutrino[29], needed to cancel possible large loop-induced neutral currents, in analogy to the proposal of the charm quark, from the GIM mechanism described in Section 1. The status of the searches for $\mu \rightarrow e\gamma$, $\mu \rightarrow eee$, $\tau \rightarrow \mu\gamma$, $\tau \rightarrow \mu\mu\mu$ and $\mu \rightarrow e$ conversions in the presence of a nucleus field as they were by the end of 2008 is shown in Figure 8. There have been significant improvements of this picture in recent years by experiments like MEG at PSI, Belle at KEKB, BaBar at SLAC and LHCb at LHC.

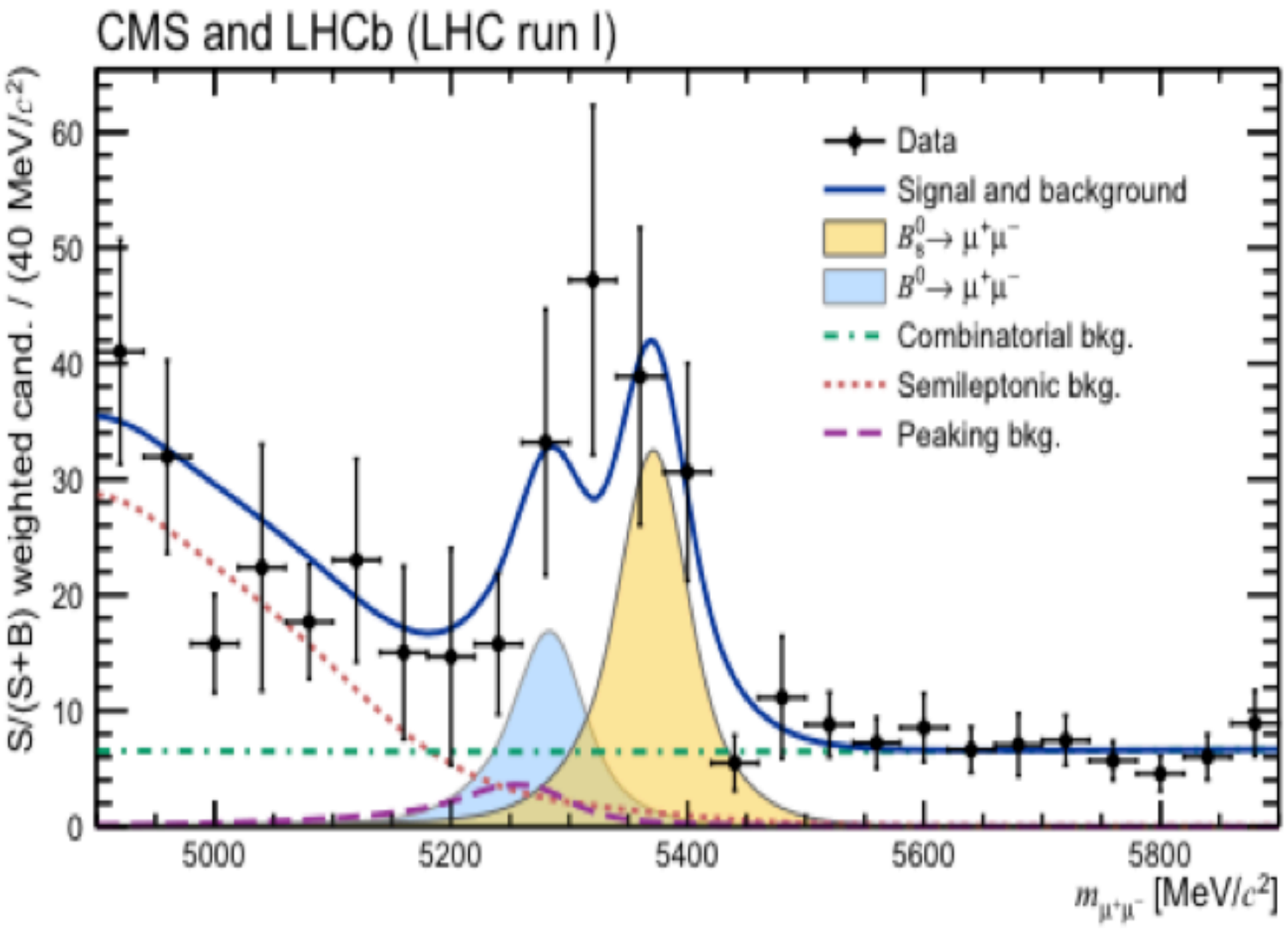

Figure 7 Weighted distribution of the dimuon invariant mass from Ref.(27). Superimposed on the data points in black are the combined fit (solid blue) and its components: the $B_s^0$ (yellow shaded) and $B_d^0$ (light-blue shaded) signal components: the combinatorial background (dash-dotted green); exclusive backgrounds (dotted pink); and the exclusive peaking backgrounds (dashed violet).

The MEG experiment collected stopped $\mu^+$ at the PSI facilities between 2009 and 2013 to search for the process $\mu^+ \rightarrow e^+\gamma$. The experimental signature consists of a monochromatic positron and photon back-to-back in the $\mu^+$ rest frame. Therefore, the experiment needs an excellent energy and tracking resolution as well as a precise measurement of the timing. The MEG collaboration has recently published[30] the results using $3.6\times10^{14}$ stopped $\mu^+$ collected up to 2011. The very few events seen in the expected signal region are compatible and slightly below the expected background, reaching an observed limit of $BR(\mu^+ \rightarrow e^+\gamma) < 5.7\times10^{-13}$ at 90% C.L. The expected sensitivity will improve when adding the rest of the data available, and should improve by a factor ten with the upgrade of the experiment.

Large samples of $\tau$ leptons are produced in high luminosity $e^+e^-$ colliders through the process $e^+e^- \rightarrow \tau^+\tau^-$. In fact, the analysis of the whole data sample collected by BaBar[31] and Belle[32], (~500-800 $fb^{-1}$ which corresponds to ~$10^9$ $\tau^+\tau^-$ pairs) shows no evidence of LFV $\tau$ decays, and, for example, improves the limits on $BR(\tau \rightarrow \mu\gamma)$ and $BR(\tau \rightarrow \mu\mu\mu)$ to a few $10^{-8}$. Moreover, the start of the LHC proton-proton collider and the enormous production of charm and beauty mesons, and subsequently of $\tau$ leptons from their decays, has opened a new window of opportunity increasing the $\tau$ production rate by five orders of magnitude w.r.t. SLAC and KEKB. The LHCb experiment has collected 3 $fb^{-1}$ of proton-proton collisions at 7-8 TeV center-of-mass energies by the end of 2012, which correspond to $\sim 10^{11}$ $\tau$ leptons. The LHCb analysis[33] has less efficiency and purity than that of the $e^+e^-$ colliders, but nevertheless reaches similar sensitivities for the decay $\tau \to \mu\mu\mu$. Already with this initial sample from LHCb the combined values as obtained by HFAG[34], improve the existing limits to: $BR(\tau \to \mu\mu\mu) < 1.2\times 10^{-8}$ and $BR(\tau \to \mu\gamma) < 5\times 10^{-8}$ at 90% C.L.

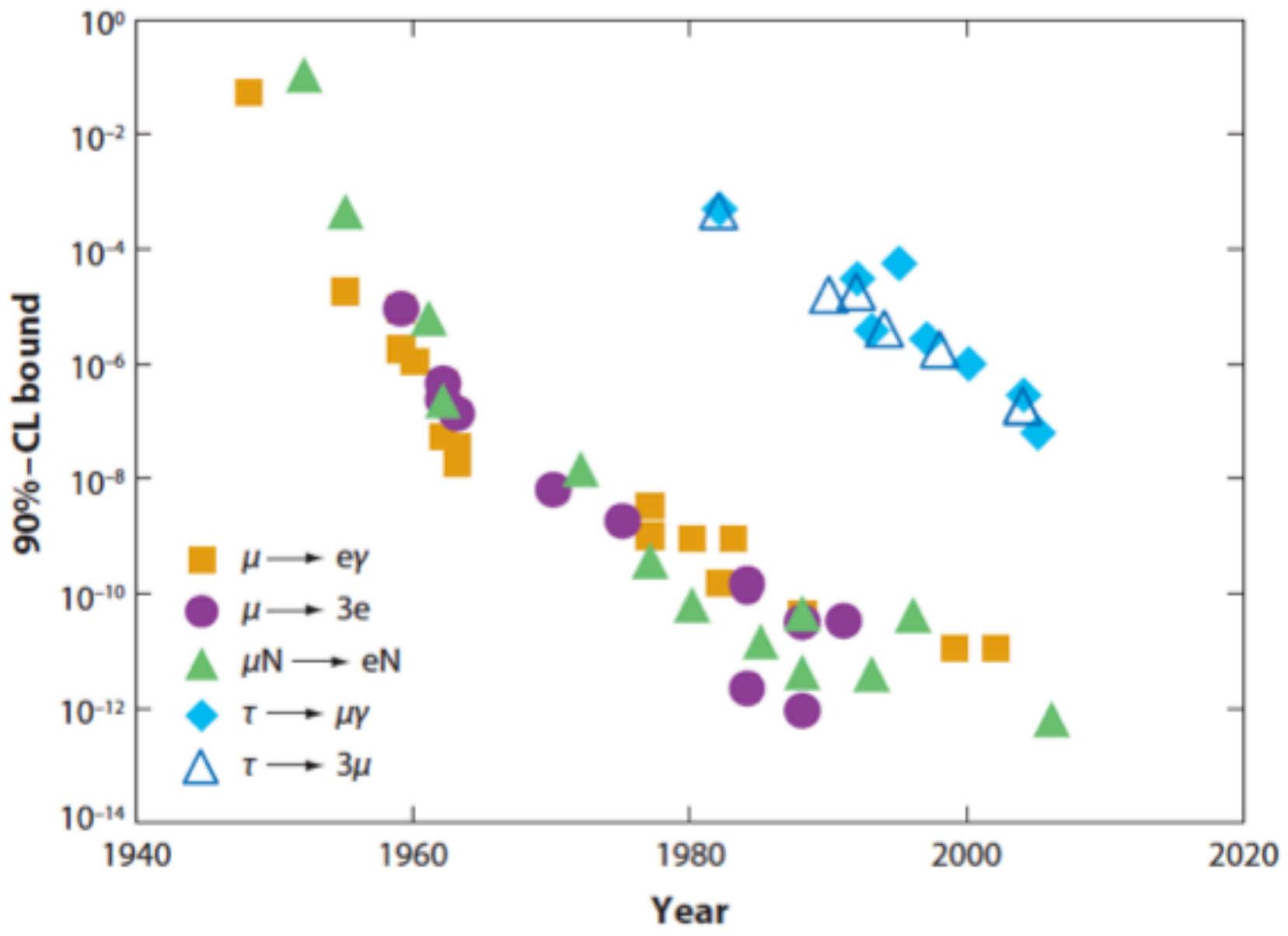

Figure 8 Historical evolution of the 90% C.L. limits on μ and τ flavour violation decays as it was in 2008 taken from Ref. (28).

Improvements in the forthcoming decade are expected from the upgrade of MEG and Belle experiments and from the data accumulated at the LHC and the upgrade of the LHC experiments. Moreover new experiments are being built at J-PARC (COMET)[35] and the Fermilab booster (Mu2e)[36] which plan to use very large samples of $\mu$ ($\sim 10^{10}$ per second) to search for $\mu \to e$ conversions in the presence of a nucleus field. They both expect to reach sensitivities below $10^{-16}$ on these processes, i.e. a four order of magnitude improvement. It is clear that Figure 8 will look very different in ten years from now.

## 5. Final remarks

The concept of symmetry has played a fundamental role in the construction of the ST. Because of these symmetries some processes are highly suppressed in the ST, but may receive important NP contributions if these symmetries are not respected by the theory that should supersede the ST. Flavour is one of these properties of the ST that cannot be explained by fundamental symmetries: why do we have three almost identical replicas of quarks and leptons, and what is the origin of their different masses? This new theory of flavour should be able to answer these questions, and naturally should include new flavour transitions. This is the reason why this chapter has been devoted to describe a few of the most interesting attempts to look for FCNC in quark and lepton decays. The field is very active today, and running experiments together with new experiments scheduled to start taking data in the next decade will provide new hope for a deeper understanding of nature.

## Acknowledgements

I would like to thank L. Maiani and G. Rolandi for their invitation to contribute to this interesting book. A. Ceccuci, S. Kettell and T. Komatsubara for their help with the text and figures in Section 3.2, and M. Kenzie for his very useful comments to the manuscript.